\documentclass[aps,preprint,preprintnumbers,amsmath,amssymb]{revtex4}
\usepackage{mathrsfs}
\usepackage{amssymb}
\usepackage{graphicx}
\usepackage{dcolumn}
\usepackage{bm}
\usepackage{color}
\usepackage[breaklinks=true,colorlinks=true,linkcolor=blue,urlcolor=blue,citecolor=blue]{hyperref}
\UseRawInputEncoding   

\usepackage{longtable}

\usepackage{multirow}
\usepackage{setspace}
\usepackage{soul}
\usepackage{ulem}
\usepackage{ragged2e}
\makeatletter
\newcommand{\rmnum}[1]{\romannumeral #1}
\newcommand{\Rmnum}[1]{\expandafter\@slowromancap\romannumeral #1@}

\begin{document}

\title{Kondo-coupled van der Waals antiferromagnet with high-mobility quasiparticles}

\author{Hai Zeng$^{1*}$, Yang Zhang$^{1*}$, Bingke Ji$^{2}$\footnote[1]{These authors contribute equally to this work}, Jiaqiang Cai$^{3}$, Shuo Zou$^{1}$, Zhuo Wang$^{1}$, Chao Dong$^{1}$, Kangjian Luo$^{1}$, Yang Yuan$^{1}$, Kai Wang$^{4}$, Jinglei Zhang$^{3}$, Chuanyin Xi$^{3}$, Junfeng Wang$^{1}$, Liang Li$^{1,5}$, Yaomin Dai$^{2}$\footnote[2]{Electronic address: ymdai@nju.edu.cn}, Jing Li$^{1}$\footnote[3]{Electronic address: jing\_li@hust.edu.cn}, and Yongkang Luo$^{1,5,6}$\footnote[4]{Electronic address:  mpzslyk@gmail.com}}

\address{$^1$Wuhan National High Magnetic Field Center and School of Physics, Huazhong University of
Science and Technology, Wuhan 430074, China;}
\address{$^2$School of Physics, Nanjing University, Nanjing 210023, China;}
\address{$^3$Anhui Key Laboratory of Low-Energy Quantum Materials and Devices, High Magnetic Field Laboratory, Hefei Institutes of Physical Science, Chinese Academy of Sciences, Hefei 230031, China;}
\address{$^4$School of Physics and Electronics, Henan University, Kaifeng 475004, China;}
\address{$^5$State Key Laboratory of Advanced Electromagnetic Technology, Huazhong University of Science and Technology, Wuhan 430074, China;}
\address{$^6$Lead contact}

\date{\today}


\maketitle

\textbf{
SUMMARY:
Two-dimensional van der Waals (vdW) materials exhibit high carrier mobility and tunability, making them suitable for low-power, high-performance electronic and spintronic applications. Incorporating narrow-band electronic correlation effects could further promote tunability, though mass renormalization may impact carrier mobility. It is therefore challenging to identify a vdW material with both high mobility and strong correlation. Herein, by a combination of optical spectroscopy and high-field quantum-oscillation measurements, we observe significant effective-mass enhancement in CeTe$_3$ at low temperature, arising from not only the band-structure modulation by antiferromagnetic ordering but also the narrow-band correlation effect. Despite the mass enhancement, the quantum mobility surprisingly \textit{increases} and reaches $\sim$2403 cm$^2$/Vs, likely benefiting from topological protection. Remarkably, these unique properties are maintained in atomically thin nanoflakes with quantum mobility enhanced to $\sim$3158 cm$^2$/Vs. Thus, CeTe$_3$ emerges as a promising Kondo-coupled vdW antiferromagnetic metal with high-mobility quasiparticles, potentially unlocking new device concepts.
}



\begin{figure*}[!htp]
	\vspace*{-5pt}
	\includegraphics[width=9.3cm]{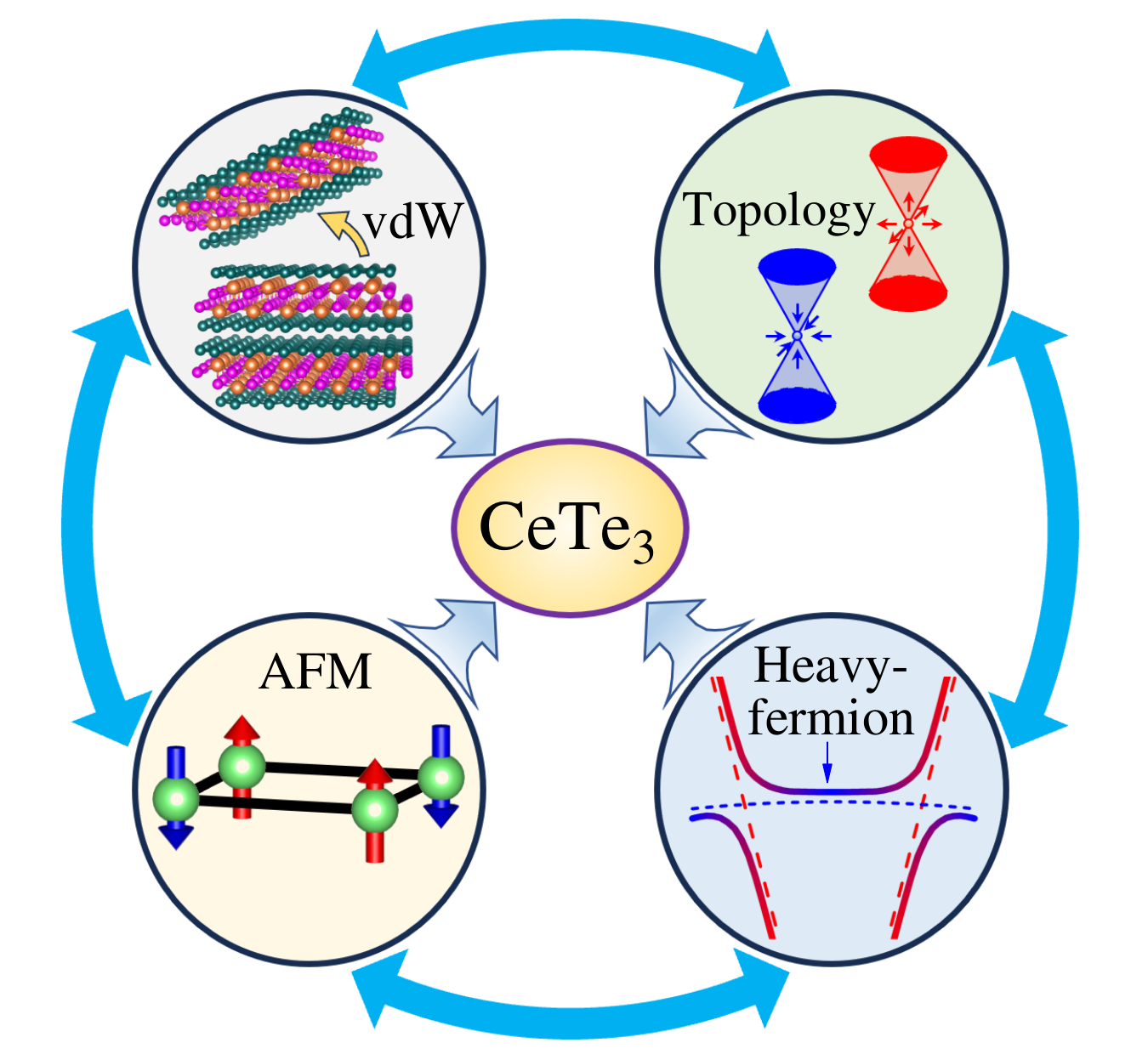}
	\label{Fig.0}
\vspace*{-5pt}
\Large
\begin{center}
    \textbf{Graphical abstract}
\end{center}
\normalsize
\end{figure*}

\clearpage

\textbf{Highlights}

$\bullet$ vdW CeTe$_3$ provides a rare example of strongly-correlated AFM metal with high mobility

$\bullet$ Complete Fermi surface topology of CeTe$_3$ is mapped out by quantum oscillations

$\bullet$ Two mechanisms for quasiparticle-mass enhancement are confirmed in CeTe$_3$

$\bullet$ High mobility, AFM order, and correlation maintained in atomically thin nanoflakes \\

\textbf{INTRODUCTION} \\

Two-dimensional (2D) van der Waals (vdW) materials have been an active research area in condensed matter physics and materials science by virtue of their unique weak interlayer coupling, high carrier mobility ($\mu$), and strong tunability, exhibiting a rich variety of quantum phenomena, such as 2D magnetism \cite{burch_2018, mak_2019}, superconductivity \cite{saito_2017, torma_2022, pantaleon_2023}, strong exciton effects \cite{mak_2018, wang_2018}, and topological states \cite{culcer_2020, zhang_2022}, bearing a great prospect in high-efficient spintronic and electronic applications. These properties underpin promising applications in next-generation spintronic and electronic technologies. To further amplify the tunability of these systems, an effective strategy is to introduce strong electronic correlations - particularly by incorporating fluctuating $f$ electrons (e.g., Ce-$4f^1$, Yb-$4f^{13}$). The strong hybridization between localized $f$ electrons and itinerant conduction ($c$) electrons (i.e., $c$-$f$ Kondo hybridization) can give rise to a renormalized narrow-band electronic structure (Figure 1A), which is highly sensitive to tuning parameters such as temperature, doping, pressure, and magnetic field. This approach opens a new frontier: vdW-layered heavy-fermion systems \cite{CeSiI-Nature}, which offer an unprecedented opportunity to integrate strong electron-electron interactions into a 2D framework, and bridge the fields of correlated electron physics and layered quantum materials, potentially unlocking new device concepts. However, as stronger correlations typically come at the cost of reduced mobility due to an increased quasiparticle effective mass ($m^*$), following $\mu = e\tau/m^*$ (where $\tau$ is the relaxation time), therefore, it is extremely challenging to tailor a high-mobility vdW material within the context of strong correlation.

\begin{figure}[!ht]
\vspace*{-0pt}
\hspace*{-0pt}
\includegraphics[width=16.5cm]{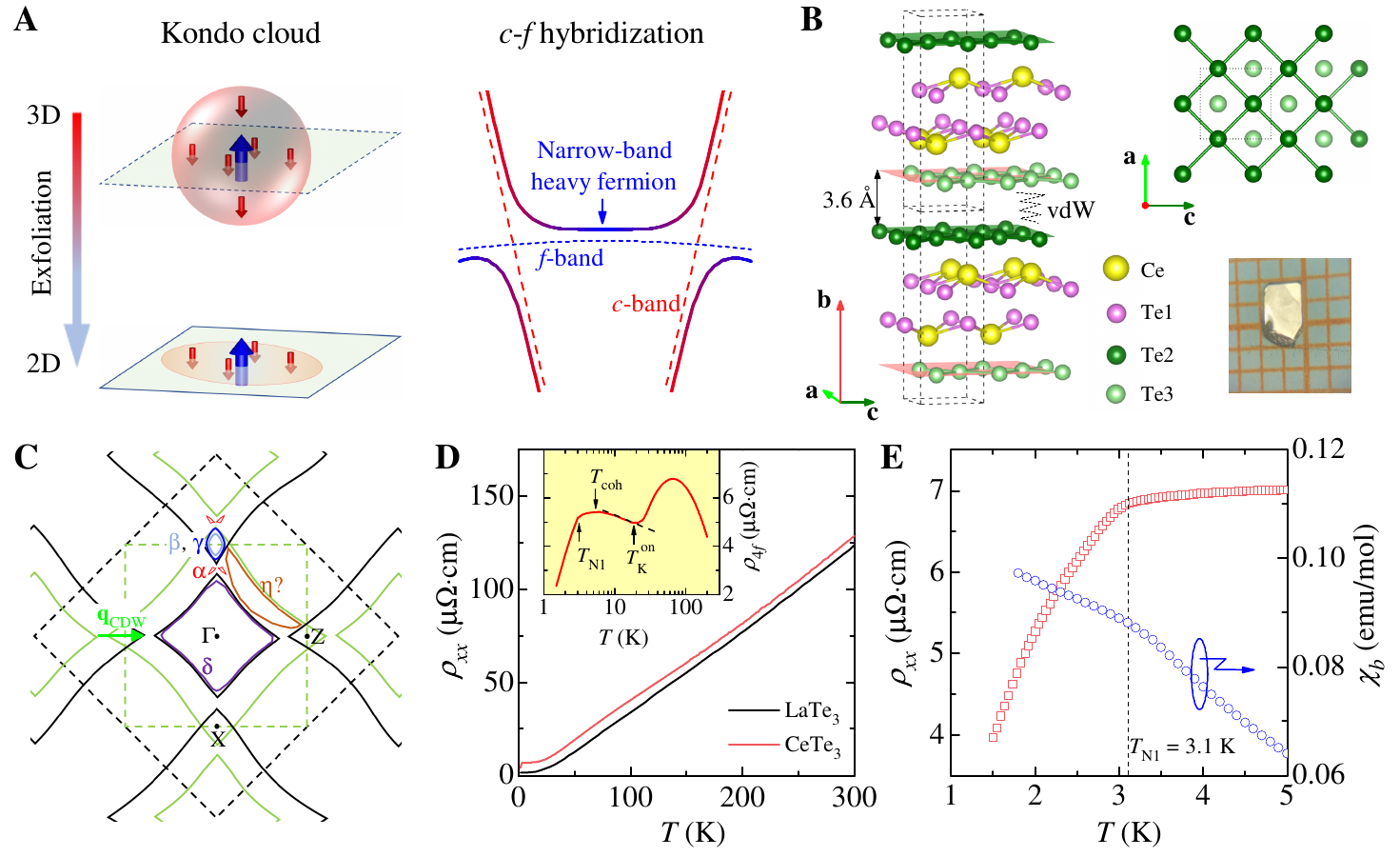}
\vspace*{-20pt}
\small
\begin{flushleft}
\justifying{
    \textbf{Figure 1. Crystal structure, Fermi surface sketch, and sample characterization.}\\ (A)
    Schematics of Kondo cloud in three-dimensional (3D) and two-dimensional (2D) systems. The blue and red arrows represent localized magnetic moments and conduction electron spins, respectively. The $c$-$f$ hybridization leads to narrow-band heavy fermions.\\ (B) Crystal structure of CeTe$_3$ and a photograph of CeTe$_3$ single crystal on millimeter-grid paper.\\ (C) The sketch of Brillion zone and Fermi surface of CeTe$_3$, proposed according to Refs.~\cite{CeTe3-FS,NdTe3-FS}. The black dashed lines and solid lines are BZ and FS of a single square net of Te atoms; the green dashed lines and solid lines are BZ and FS of 3D structured CeTe$_3$; the FS is reconstructed due to the back folding and CDW-induced nesting effect, which finally leads to the $\alpha$ (red), $\beta$ (cyan), $\gamma$ (blue), $\delta$ (purple) pockets. A guessed $\eta$ (orange) pocket is also shown.\\ (D) Temperature-dependent resistivity of bulk CeTe$_3$ (pink) and LaTe$_3$ (black). The inset shows $\rho_{4f}$ ($\equiv\rho_{\text{Ce}}-\rho_{\text{La}}$), the resistivity contribution from scattering by $4f$ electrons. The dashed line signifies the $-\log T$ behavior due to incoherent Kondo scattering.\\ (E) Low-temperature resistivity and magnetic susceptibility both show the AFM1 transition at $T_\text{N1}=3.1$ K. The AFM2 transition at $T_\text{N2}$ is beyond the scope of our measurements.
    }
\end{flushleft}
\normalsize
\label{Fig1}
\end{figure}

The rare-earth tritelluride family $R$Te$_3$ ($R$ = La-Tm, Y) has recently garnered extensive attention for coexistence of multiple charge-density-wave (CDW) orders \cite{RTe3-breakdown,RTe3-OS,RTe3-CDW,CeTe3-STM,CeTe3-CDW,CeTe3-gap}, pressure- and intercalation-induced superconductivity \cite{CeTe3-Pressure,RTen-Pd inter}, rich magnetic states \cite{Magnetic-TbTe3,Magnetic-RTe3,Magnetic-CeTe3}, and notably high mobility \cite{GdTe3-Sci adv,LaTe3-interplay}. Among them, CeTe$_3$ is of particular interest in that the Ce-$4f^1$ configuration makes it a candidate 2D Kondo lattice. Yet, the role of electronic correlation in CeTe$_3$ remains poorly understood. Furthermore, unlike LaTe$_3$ and other lanthanide counterparts - where quantum-oscillation measurements have identified multiple well-resolved Fermi pockets - the Fermi surface (FS) topology of CeTe$_3$ is less clear \cite{CeTe3-QO1,CeTe3-QO2,CeTe3-QO3,RTe3-breakdown,Smith2024-CeTe3}. This gap in knowledge obscures a full understanding of its quantum transport properties and the role of correlation effects.

In this work, we perform a comprehensive study of bulk CeTe$_3$ via optical spectroscopy and high-field quantum oscillations. Besides the known low-frequency Fermi pocket ($\alpha$), we identify four additional pockets ($\beta$, $\gamma$, $\eta$, and $\delta$) with larger cross-sections, completing the bulk FS topology. A significant enhancement of carrier effective mass is observed around the onset of Kondo effect ($T_\text{K}^\text{on}\sim 17$ K), lending further support to the existence of moderately heavy fermions in this 2D Kondo-lattice compound.  A second mass enhancement occurs at the antiferromagnetic (AFM) transition, accompanied by an unexpected increase in quantum mobility and a 0-to-$\pi$ Berry phase shift, suggesting a topological reconstruction of the Fermi surface. We further extend our investigation to the atomically thin limit. Strikingly, all the key features are retained in few-monolayer nanoflakes. These results establish CeTe$_3$ as a rare vdW system that hosts both strong electronic correlation and high charge mobility even at the 2D limit, and thus lay a foundation for designing miniaturized spintronic devices leveraging strong electronic correlation and robust magnetic order. \\

\textbf{RESULTS} \\

\textbf{Crystal structure and sample characterization}

CeTe$_3$ crystallizes in an orthorhombic structure with the space group Cmcm (No. 63), as shown in Figure 1B. It consists of Ce-Te slabs sandwiched by Te-bilayer square nets stacking along the $\mathbf{b}$-axis (perpendicular to the Te nets). The adjacent Te-Te layers are weakly coupled by vdW force, and can be mechanically exfoliated into nanoflakes. The contributions to the FS are primarily from $p_x$ and $p_z$ orbitals of Te2 and Te3 in the 2D square nets, rendering a 2D characteristic (Figure 1C) \cite{CeTe3-CPB}.

The temperature-dependent resistivity of bulk CeTe$_3$ is displayed in Figure 1D, where the results of its non-$4f$ reference, LaTe$_3$, are also provided for comparison. In general, CeTe$_3$ exhibits a good metallic behavior with a large residual resistivity ratio [RRR $\equiv \rho(300 \text{K})/\rho(2 \text{K}) = 40$], indicating the high quality of the single crystal. The scatterings contributed from $4f$ electrons, $\rho_{4f}$, can be separated out by subtracting the resistivity of LaTe$_3$ (inset to Figure 1D). Around 100 K, a noticeable bump is observed in $\rho_{4f}(T)$, which should be ascribed to the scatterings between crystalline-electric-field (CEF) split doublets, as is the case in most Ce-contained compounds \cite{LuoY-CeNi2As2,ChenK-CeCo2Ga8}. Below $\sim 17$ K, Kondo effect sets in due to the $c$-$f$ hybridization. Initially, the size of Kondo clouds (cf. Figure 1A) is small, so the Kondo scatterings are incoherent, characterized by the $-\log{T}$ law in $\rho_{4f}(T)$. Upon cooling, the Kondo coupling strengthens, and Kondo clouds communicate with each other, yielding the development of Kondo coherence that is manifested by  a broad maximum in $\rho_{4f}(T)$ near $T_\text{coh} = 5.5$ K. A clear inflection is seen at $T_\text{N1}=3.1$ K, due to the reduction of spin scattering in the AFM ordered state \cite{CeTe3-QO3}. The transition at $T_\text{N1}$ (denoted by AFM1) is also manifested in the magnetic susceptibility (Figure 1E). We are not able to see the second AFM transition at $T_\text{N2}=1.3$ K (AFM2) \cite{Magnetic-CeTe3} that is lower than the base temperature (1.5 K) of our measurements. It should be mentioned that previous specific heat measurements revealed a large Sommerfeld coefficient $\sim 400$ mJ/mol K$^2$ in the paramagnetic state \cite{Magnetic-CeTe3}, indicative of mass enhancement due to electronic correlation effect. All these suggest CeTe$_3$ as a candidate vdW AFM dense Kondo lattice. \\

\textbf{Optical spectroscopy of bulk CeTe$_3$}

\begin{figure*}[!htp]
\vspace*{0pt}
\hspace*{0pt}
\includegraphics[width=16.5cm]{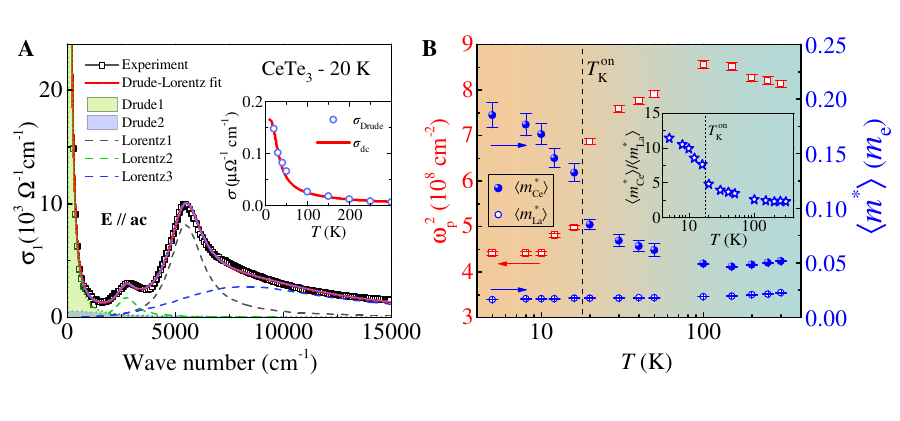}
\vspace*{-30pt}
\small
\begin{flushleft}
\justifying{
    \textbf{Figure 2. Optical conductivity of CeTe$_3$.}\\ (A) Drude-Lorentz fitting of $\sigma_1(\omega)$, taking $T=20$ K as an example. The three Lorentz components are depicted by dashed lines, whereas the two Drude components are marked in color-filled areas. The inset shows the temperature dependence of DC conductivity $\sigma_\text{dc}$ and the total contribution of Drude terms, $\sigma_\text{Drude}$, in the limit $\omega \rightarrow 0$.\\ (B) Left, temperature dependence of $\omega_\text{p}^2\equiv\omega_\text{p1}^2+\omega_\text{p2}^2$; Right, temperature dependent average effective mass of charge carriers, $\langle m^{*} \rangle$. For comparison, the $\langle m^{*} \rangle$ of LaTe$_3$ is also shown. The inset shows the mass-enhancement ratio, $\langle m^{*}_\text{Ce} \rangle/\langle m^{*}_\text{La} \rangle$.
    }
\end{flushleft}
\normalsize
\label{Fig2}
\end{figure*}

To further look into the electronic correlation effect in CeTe$_3$, we measured the optical spectroscopy at various temperatures [see Figure S2 in \textbf{Supplemental Information} (\textbf{SI}) for raw data]. In the real part of the optical conductivity, $\sigma_1(\omega)$, the features for the CDW gaps reported previously \cite{sacchetti2006,sacchetti2007,CeTe3-OS}, e.g. the pronounced peak structure around 5000 cm$^{-1}$ for all the measured temperatures and a second smaller peak centered near 2800 cm$^{-1}$ present only for below 200 K, are well reproduced in our data. Here we mainly focus on the mass renormalization, and this can be obtained by the quantitative analysis of the optical conductivity. As shown in Figure~2A, $\sigma_1(\omega)$ can be well fitted to the Drude-Lorentz dispersion model \cite{Dressel-2002,Dai-La3Ni2O7},
\begin{equation}
    \sigma_1(\omega) = \frac{2\pi}{Z_0} \left[\sum_k  \frac{\omega_{\text{p},k}^2}{\tau_k(\omega^2 + \tau_k^{-2})} + \sum_i \frac{\gamma_i \omega^2 \Omega_{i}^2}{(\omega_{i}^2 - \omega^2)^2 + \gamma_i^2 \omega^2} \right],
    \label{Eq1}
\end{equation}
where $Z_0 = 376.73~\Omega$ is vacuum impedance. The first term of Eq.~(\ref{Eq1}) represents the sum of the Drude responses for free charge carriers (intraband transition), with $\omega_{\text{p},k}$ and $1/\tau_k$ being the $k$-th component of plasma frequency and quasiparticle scattering rate, respectively. The second term of Eq.~(\ref{Eq1}) is the sum of the Lorentz responses for localized electrons (interband transition) in which $\omega_{i}$, $\gamma_i$, and $\Omega_{i}$ stand for the resonance frequency, damping, and strength of the $i$-th oscillator, respectively. The fit employs two Drude terms, manifested respectively by the areas in green and blue in Figure 2A. The total Drude contribution in the limit $\omega \to 0$ matches well with the $\sigma_\text{dc}(T)$ profile, where $\sigma_\text{dc} \equiv 1/\rho$ is the DC conductivity (inset of Figure 2A). 
Other key parameters obtained from this analysis are $\omega_\text{p1}^2$ and $\omega_\text{p2}^2$, the sum of which (denoted as $\omega_\text{p}^2$) is displayed in Figure 2B as a function of temperature. One salient feature is that $\omega_\text{p}^2$ is essentially constant above 30 K, but decreases substantially below $T_\text{K}^\text{on} \sim 17$ K, manifesting an enhancement of quasiparticle effective mass due to Kondo hybridization. To see this more clearly, we estimate the average effective mass via $\langle m^* \rangle = \frac{Z_0 n e^2}{2\pi \omega_\text{p}^2}$, where $n$ is the total carrier concentration that can be derived by Hall effect measurements [Figures S5 and S6 in \textbf{SI}]. A tiny decrease in $\langle m^* \rangle$ is observed from 300 K to 150 K (Figure 2B), likely due to the opening of a second CDW gap around 200 K \cite{CeTe3-OS}. Owing to the multiple CDW gaps, on the whole, $\langle m^* \rangle$ is small for all temperatures below 300 K. However, the \textit{remnant} quasiparticles still exhibit prominent mass renormalization. Below 100 K, $\langle m^* \rangle$ gradually increases, while a substantial enhancement is observed around $T_\text{K}^\text{on}$. Recently, there has been accumulating evidence that fluctuating $c$-$f$ hybridization induced narrow bands start to appear near $T_\text{K}^\text{on}$ in prototypical Kondo-lattice compounds (e.g. CeCoIn$_5$, CeCo$_2$Ga$_8$, etc), as exemplified by a series of angle-resolved photoemission spectroscopy (ARPES) \cite{ChenQY-PRB2017,KoitzschA-PRB2008}, ultrafast optical spectroscopy \cite{LiuY-PRL2020}, and planar Hall effect \cite{ZouS-CeCo2Ga8PHE2025} experiments. It, therefore, is reasonable to ascribe the $\langle m^* \rangle$ enhancement of CeTe$_3$ around $T_\text{K}^\text{on}$ to the Kondo coupling effect. At the lowest temperature of our optical measurements, 5 K, the ratio $\langle m^{*} \rangle/\langle m^{*}_\text{300K} \rangle \approx 3.6$. In order to further demonstrate the effective mass enhancement, we also performed the same optical measurements on the non-$4f$ reference LaTe$_3$, seeing Figure S4. The electronic correlation effect in CeTe$_3$ can be directly visualized through the ratio $\langle m_\text{Ce}^{*} \rangle/\langle m_\text{La}^{*} \rangle$ that rises rapidly below $T_\text{K}^\text{on}$ and reaches a large value of $\sim 11.4$ at 5 K (inset to Figure 2B). Though the factors of $\langle m^{*} \rangle/\langle m^{*}_\text{300K} \rangle$ and $\langle m_\text{Ce}^{*} \rangle/\langle m_\text{La}^{*} \rangle$ are moderate when compared with classic heavy-fermion metals, they are much larger than traditional gate-tuned 2D systems such as GaAs/AlGaAs heterostructure \cite{TanY-GaAsPRL2005}, bilayer graphene \cite{ZouK-BGraphenePRB2011}, etc. All these place CeTe$_3$ in a regime with moderate electronic correlation. Note that earlier ARPES measurements on CeTe$_3$ also revealed a weakly dispersive quasiparticle band near the Fermi level at low temperature \cite{CeTe3-CPB}, in consistency with the mass enhancement due to Kondo hybridization.  \\

\textbf{Quantum oscillations of bulk CeTe$_3$}

Before running into our quantum oscillation results, it is necessary to put forward a qualitative sketch of Fermi-surface structure (Figure 1C). Due to the larger unit cell in the 3D crystal structure compared with the single square nets, the Brillouin zone (BZ) is reduced (from black dashed square to green dashed square). Correspondingly, the original FS (black solid lines) is ``folded" at the border of the reduced BZ. The nesting of the FS in the presence of CDW order not only removes FS elements but also creates new fractions \cite{CeTe3-FS,NdTe3-FS}, which eventually results in multiple FS pockets. Although multiple Fermi sheets were confirmed by quantum oscillations in other members of $R$Te$_3$ \cite{RTe3-breakdown}, no such results have been reported in CeTe$_3$ \cite{CeTe3-QO1,CeTe3-QO2,CeTe3-QO3,Smith2024-CeTe3}. Since enhanced effective mass typically means lower Fermi velocity, shorter transport mean free path and smaller carrier mobility, presumably, high magnetic field will be required to map out its FS topology by quantum oscillations.

\begin{figure*}[!htp]
\vspace*{10pt}
\hspace*{-0pt}
\includegraphics[width=16.5cm]{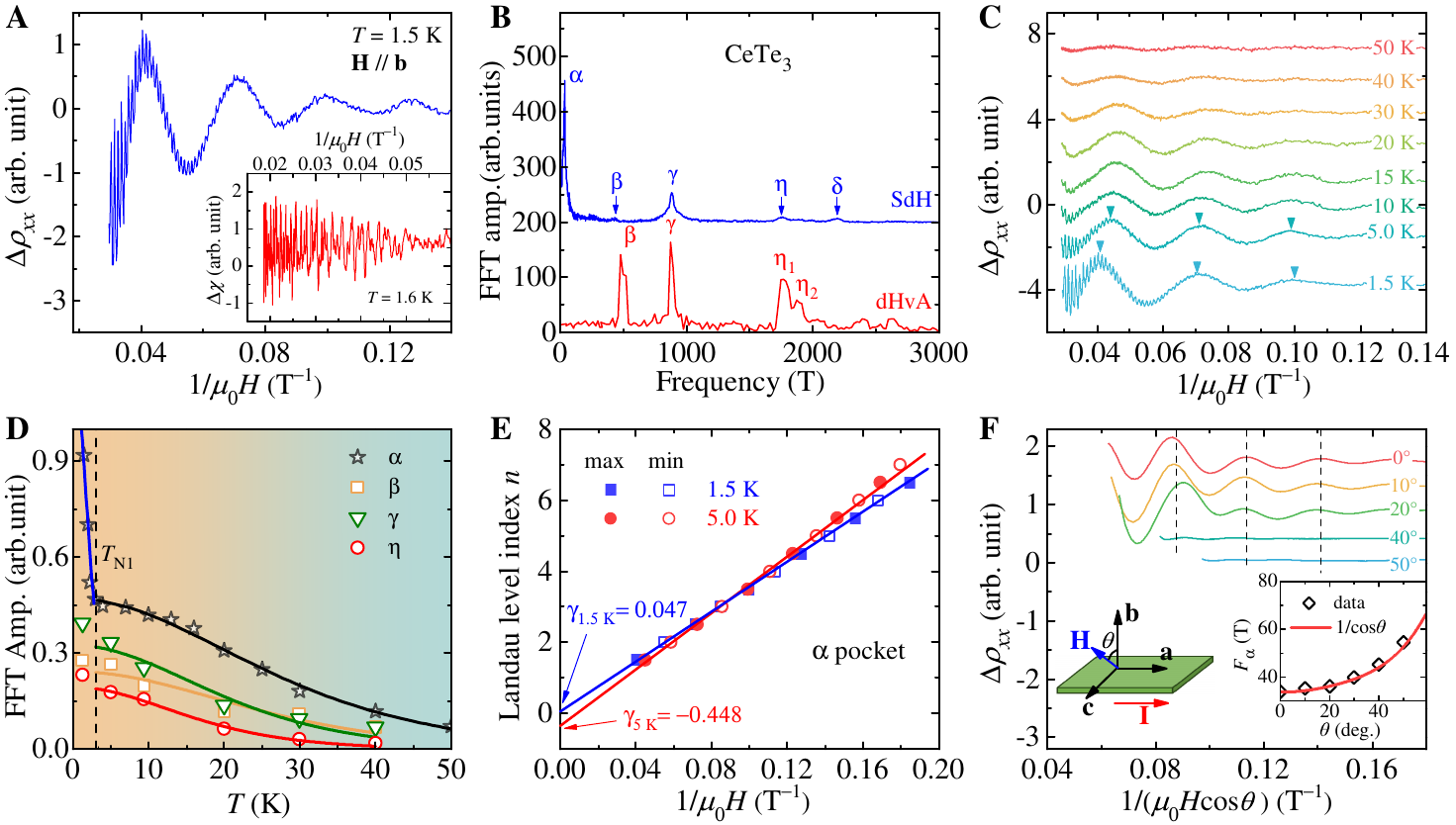}
\vspace*{-20pt}
\small
\begin{flushleft}
\justifying{
    \textbf{Figure 3. Quantum oscillations of CeTe$_3$ bulk sample.}\\ (A) SdH oscillatory component at 1.5 K. The inset shows dHvA oscillations at 1.6 K.\\ (B) FFT spectra of SdH and dHvA oscillations, which reveal multiple frequencies, $F_{\alpha-\eta}$.\\ (C) SdH oscillations at various temperatures.\\ (D) FFT amplitudes as functions of $T$, and their LK fits (solid lines). Data of $\alpha$ are from SdH (4 - 16 T), while the remaining are from dHvA (20 - 54 T). An abrupt upturn is clearly seen at $T_\text{N1}$, especially for the $\alpha$ pocket, which gives rise to different effective masses for above and below $T_\text{N1}$.\\ (E) Landau fan index diagram of the $\alpha$ oscillation at 5 K (red) and 1.5 K (blue). The solid lines represent the linear fittings that yield the intercepts $-0.448$ and $0.047$, respectively.\\ (F) SdH oscillations at 1.5 K under rotational field within $\mathbf{bc}$ plane. $F_{\alpha}$ conforms to the $1/\cos{\theta}$ law, demonstrating 2D nature of this material.
    }
\end{flushleft}
\normalsize
\label{Fig3}
\end{figure*}

Here, we exploit both Shubnikov-de Haas (SdH) and de Haas-van Alphen (dHvA) effects under high magnetic field up to 34 T and 54 T, respectively. Clear SdH oscillations are visible in both longitudinal ($\rho_{xx}$) and transverse ($\rho_{yx}$) channels, seeing Figure S5. The oscillatory part $\Delta\rho_{xx}$ can be separated out by subtracting the polynomial background and is plot versus $1/\mu_0H$ in Figure 3A. The SdH effect in quasi-2D systems is usually described by the Lifshitz-Kosevich (LK) formula \cite{DShoenberg-MagOscillation,LuoY-NbAsSdH},
\begin{equation}
    \Delta \rho (\mu_0H, T) \propto \frac{\lambda T}{\sinh(\lambda T)} e^{-\lambda T_\text{D}} \cos \left(\frac{2 \pi F}{\mu_0H} - \pi + \Phi_\text{B}  \right),
    \label{eq:equation_label}
\end{equation}
where $\lambda=(2\pi^2k_Bm^*)/(\hbar e \mu_0 H)$, $T_\text{D}$ is Dingle temperature, and $\Phi_\text{B}$ is Berry phase. Multiple oscillation frequencies are discernible in Fast Fourier Transform (FFT) spectrum of $\Delta\rho_{xx}(1/\mu_0H)$, seeing Figure 3B. Besides the previously reported predominant oscillation frequency $F_\alpha = 36$ T that shows up at relatively low field ($<8$ T) \cite{CeTe3-QO1,CeTe3-QO2,CeTe3-QO3}, our high-field SdH measurements unveil four additional frequencies, $F_\beta = 478$ T, $F_\gamma = 884$ T, $F_\eta = 1760$ T, and $F_\delta = 2200$ T. Quantum oscillations were also observed in magnetization measurements, viz dHvA effect, seeing Figure S1C and the inset to Figure 3A. It seems the dHvA effect is more sensitive to those high-frequency oscillations (e.g., $F_\beta$), whereas the smallest $F_{\alpha}$ is hardly seen. In particular, frequency splitting of $F_{\eta}$ is also distinguishable in dHvA, which is likely a consequence of weak FS warping and bilayer splitting \cite{RTe3-breakdown}. Table 1 summarizes the information of all the observed quantum oscillations. It is well known that the oscillation frequency is related to the extrema cross-sectional area $A_{F}$ of FS (perpendicular to the applied field) through the Onsager relation $F=(\hbar/2\pi e)A_{F}$ \cite{Shoenberg-MagneticOscillations}. The observed oscillation frequencies correspond to $0.16\%$, $2.16\%$, $3.99\%$, $7.94\%$ and $9.93\%$ of the entire BZ, respectively. According to previous ARPES studies \cite{CeTe3-FS,NdTe3-FS}, $F_{\alpha}$ arises from a small pocket in the butterfly-shaped FS wing near the X point (Figure 1C); $F_{\beta}$ and $F_{\gamma}$ can be ascribed to the elliptical pockets around the X point and are consequences of the Fermi surface reconstruction due to the CDW transition; the largest $F_\delta$ likely corresponds to the square pocket in the center of BZ. The exact origin of $F_{\eta}$ is still unclear based on ARPES results \cite{NdTe3-FS}; here we tentatively assign it to the section nearby the $\delta$. Another possibility for $F_\eta$ is the second harmonics of the $\gamma$ pocket. ARPES  with improved resolution will be needed to clarify this issue. It should also be pointed out that the FS structure is slightly modulated across the AFM1 transition, cf. Table 1. We will revisit this issue later.

\begin{table*}[!htp]
\caption{\label{TabS1}Quantum oscillation and transport parameters of each Fermi pocket of CeTe$_3$ for $\mathbf{H}\parallel\mathbf{b}$. PM = Paramagnetic; AFM = Antiferromagnetic.}
\begin{ruledtabular}
\begin{tabular}{cccccc}
Sample                 & Pocket     & Region   & $F$ (T)$^{\dag}$ &  $m^*~(m_\text{e})$  & $\mu_\text{q}$ (cm$^2$/Vs)                \\ \hline
\multirow{6}{*}{Bulk}  & \multirow{2}{*}{$\alpha$} & AFM (1.5 K) & 34      &  0.46(4)   &    2403      \\
                       &                           & PM (5 K)    &    36     &    0.035(1)      &   1883       \\
                       & $\beta$                   & PM (5 K)    & 478     &  0.22(4)   &  -        \\
                       & $\gamma$                  & PM (5 K)    & 884     &  0.27(3)   &  -        \\
                       & $\eta$                    & PM (5 K)    & 1760    &  0.34(2)   &  -        \\
                       & $\delta$                  & PM (5 K)    & 2200    &   -      &  -        \\ \hline
\multirow{5}{*}{D1}    & \multirow{2}{*}{$\alpha$} & AFM (1.5 K) &  37     & 0.70(7)    &    3158      \\
                       &                           & PM (5 K)    &     39    &    0.037(1)      &    2155      \\
                       & $\beta$                   & PM (5 K)    & 531     &   -      &  -        \\
                       & $\gamma$                  & PM (5 K)    & 846     &   -      &  -        \\
                       & $\eta$                    & PM (5 K)    & 1748    &   -      &  -        \\
                       & $\delta$                  & PM (5 K)    & 2155    &   -      &  -        \\
\end{tabular}
\end{ruledtabular}
\small
\vspace*{-10pt}
\begin{flushleft}
\justifying{
$^{\dag}$ For $\beta$, $\gamma$, $\eta$ and $\delta$ pockets, it is found that the AFM1 transition has little impact on their quantum oscillation frequencies.
}
\end{flushleft}
\normalsize
\end{table*}

Remarkably, the SdH quantum oscillations of CeTe$_3$ persist to higher than 50 K (Figure 3C), indicative of high mobility of this correlated vdW material. According to the LK formula Eq.~(2), the FFT amplitude decays with increasing temperature following a law of $\lambda T/\sinh{\lambda T}$, and the fitting of the experimental data to this formula shall lead to the effective mass $m^*$. A major feature here, however, is that the temperature dependence of FFT amplitude exhibits a distinctive inflection around $T_\text{N1}$, and this is particularly clear for $F_{\alpha}$ [Figure 3D]. Such kind of non-LK temperature dependence has been observed in a handful of materials including mercury \cite{ElliottM-JPFMP1980}, fractional quantum Hall systems \cite{TsuiD-PRL1982} and SmB$_6$ \cite{TanB-Science2015} with different mechanisms. In CeTe$_3$, since the deviation is coincident with the AFM1 transition, it is reasonable to ascribe it to the Fermi surface modulation by magnetic ordering, as mentioned by Higashihara \textit{et al} \cite{CeTe3-QO3}. Separate fittings reveal that the quasiparticle effective mass changes from $m^*_{\alpha}=0.035(1)~m_\text{e}$ for above $T_\text{N1}$ to $m^*_{\alpha}=0.46(4)~m_\text{e}$ for below. The Dingle temperature can then be deduced from the LK analysis, $T_\text{D}=1.90$ K (below $T_\text{N1}$) and $T_\text{D}=32.61$ K (above $T_\text{N1}$), and this in turn yields the quantum lifetime $\tau_\text{q}=\hbar/2\pi k_\text{B} T_\text{D}$ and quantum mobility $\mu_\text{q}=e\tau_\text{q}/m^*$. It is worthwhile to mention that the enhancement of quasiparticle effective mass below $T_\text{N1}$ does not cause a reduction of quantum mobility, but rather, $\mu_\text{q}$ is increased by $27.6\%$, and meanwhile, $\tau_\text{q}$ is enlarged by about 17 times. Such an enhancement in $\mu_\text{q}$ and $\tau_\text{q}$ reminds us that the AFM transition at $T_\text{N1}$ may be also accompanied with a topological transition. For this purpose, a Landau fan plot is constructed for the $\alpha$ FS, and the results of 1.5 K and 5.0 K are compared in Figure 3E. Our high-field SdH measurements allow determining the Berry phase accurately. For the $\alpha$ pocket, the application of 34 T magnetic field confines the carriers below the $n=2$ Landau level.
According to Lifshitz-Onsager quantization rule for 2D system \cite{LO-QO}: $A_F(\hbar/2\pi e \mu_0 H) = n+1/2-\Phi_\text{B}/2\pi$, the Berry phase $\Phi_\text{B}$ can be obtained from the intercept when linearly extrapolating the Lanau level index to the limit of $1/\mu_0H \rightarrow 0$. The derived intercepts are 0.047 and $-0.448$ for 1.5 and 5 K, meaning the corresponding Berry phase $\Phi_\text{B}$ is close to $\pi$ and 0, respectively. Additional temperature dependent Berry phase is shown in Figure S8. The nontrivial Berry phase at 1.5 K strongly suggests that CeTe$_3$ may be a kind of antiferromagnetic topological semimetal due to which the backward scattering rate is greatly suppressed by the topological protection \cite{LiangT-Cd3As2}. However, we must admit here that right now it is hard for us to further prove this by first-principles calculations, as the exact magnetic structure below $T_\text{N1}$ remains unclear to date; ARPES measurements below $T_\text{N1}$ are also challenging.

Before closing this section, we shall briefly elaborate the discrepancy in the extent of mass renormalization revealed through quantum oscillation, optical conductivity, and specific heat. Quantum oscillation is FS-dependent and mobility-sensitive, primarily probing the $m^*$ of specific light pockets, for instance $m_\alpha^*$ in our case. Optical conductivity assesses the effective mass of conduction bands through the total Drude weight [cf. the first term of Eq.~(1)], meaning that the derived effective mass represents the average of all free charge carriers, denoted as $\langle m^* \rangle$. Specific heat, on the other hand, measures the density of states from all the electronic contributions including both conduction and localized electrons. Notably, localized electrons can also enter the specific heat via magnetic entropy (e.g. magnetic transitions, spin fluctuations, Kondo hybridization, Schottky anomalies, and so on) which usually is significant for Ce-$4f$ in moderate Kondo lattice, as is the case in CeTe$_3$ \cite{Magnetic-CeTe3}. Although these different techniques yield differing values for effective mass or Sommerfeld coefficient, they consistently point to the presence of considerable electronic correlation effect in CeTe$_3$ at low temperature. \\

\textbf{2D feature and quantum transport of nanoflakes}

The results of angular dependent SdH measurements are shown in Figure 3F. As the applied field is tilted away from $\mathbf{b}$ towards $\mathbf{c}$, the dominant oscillation $F_{\alpha}$ conforms well to the $1/\cos{\theta}$ law, cf the inset to Figure 3F. This provides additional evidence for the 2D character of CeTe$_3$. The weak vdW interlayer coupling in CeTe$_3$ enables us to fabricate atomically thin films by mechanical exfoliation. Nanoflakes with thickness $\sim$38.2, 25.6 and 7.6 nm (labled by D1, D2, an D3) were successfully prepared, corresponding to 27, 18 and 6 monolayers (Figure 4A), respectively. As for the quantum transport properties of these nanoflake samples, two prominent features can be identified:

\begin{figure*}[!htp]
\vspace*{-0pt}
\hspace*{-0pt}
\includegraphics[width=16.5cm]{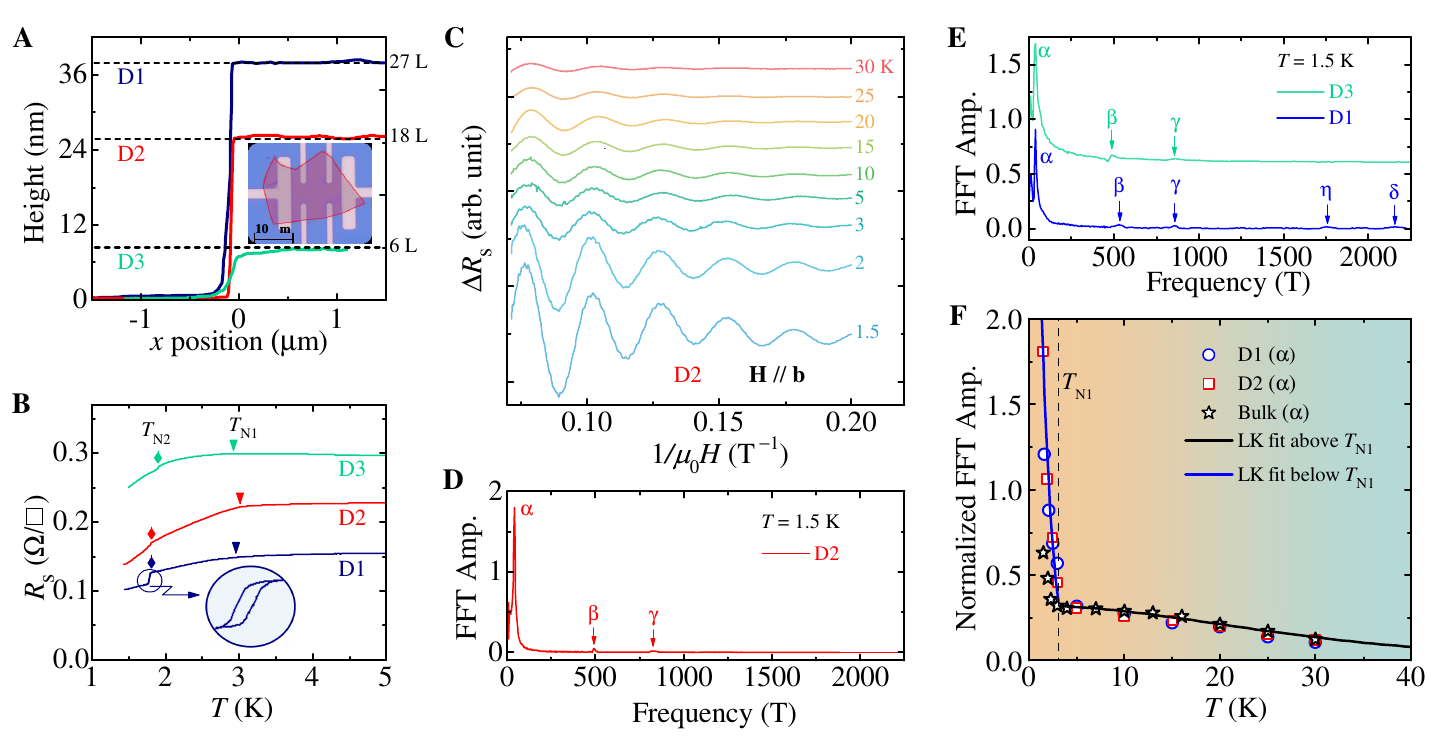}
\vspace*{-20pt}
\small
\begin{flushleft}
\justifying{
    \textbf{Figure 4. Quantum transport of nanoflaked CeTe$_3$ samples.}\\ (A) Cross-sectional height profiles of atomic force microscopy images, displaying the thickness of D1 (38.2 nm), D2 (25.6 nm), and D3 (7.6 nm). The thickness of the flakes is converted to the number of monolayers and displayed on the right axis. The inset shows an atomic force microscopy image of D2.\\ (B) Temperature-dependent sheet resistance for D1, D2, and D3 flakes, showing the two sequential AFM transitions at $T_\text{N1}$ and $T_\text{N2}$.\\ (C) SdH oscillations of D2 at various temperatures.\\ (D) FFT spectrum of D2 at 1.5 K.\\ (E) FFT spectra of D1 and D3 at 1.5 K.\\ (F) Temperature dependence of normalized FFT amplitudes of the $\alpha$ oscillation; comparison among nanoflakes D1, D2 and bulk sample. The black and blue lines represent the LK fits above and below $T_\text{N1}$, respectively. Note that the experimental data are normalized at 5 K.
    }
\end{flushleft}
\normalsize
\label{Fig4}
\end{figure*}

(\rmnum{1}) The AFM transitions survive even in the thinnest 6L nanoflake, seeing Figure 4B. The critical temperature $T_\text{N1}$ in D1-3 are only slightly decreased with respect to the bulk, although the signature of the transition in the sheet resistance $R_\text{s}(T)$ seems much washed-out. To our surprise, the AFM2 transition at $T_\text{N2}$ is promoted by exfoliation. In bulk samples, $T_\text{N2}$ was determined by a $\lambda$-shaped peak in specific heat around 1.3 K \cite{Magnetic-CeTe3}. The value now shifts to $\sim 1.8$ K in nanoflakes, as manifested by an abrupt discontinuity in $R_\text{s}(T)$ and a hysteresis loop that is most obvious in D1 (inset to Figure 4B). In other words, the AFM2 transition now becomes 1st-order-like in nature. The mechanism for these modifications remains mysterious for us. One possibility is due to a strong coupling between AFM2 and CDW orders, the latter of which can be promoted by exfoliation as a consequence of FS instability in lower dimensionality \cite{GdTe3-Sci adv}. Another possibility is that, because the AFM1 transition is softened, as a compensation, the magnetic entropy release at AFM2 is enlarged; the abundance of spin fluctuations near $T_\text{N2}$ may drive the transition into first-order \cite{ManoschekJ-MnSiPRB2013}. 
Additional experiments such as neutron scattering are required to further clarify this issue.

(\rmnum{2}) The high mobility of the bulk CeTe$_3$ is also inherited in nanoflakes, as indicated by the robust SdH oscillations persisting above 50 K (Figures 4C and S7). The FFT of D1 manifests that all the five oscillation frequencies are observable (Figure 4E) and their values are close to those in the bulk sample, seeing Table 1. D2 and D3 also exhibit pronounced quantum oscillations, but the signals for $\eta$- and $\delta$-pockets are less discernible. To make a good comparison of the effective mass between nanoflake and bulk samples, we normalize their FFT amplitudes of the $\alpha$ pocket at 5 K (Figure 4F). In the paramagnetic phase, all the curves roughly fall onto a single line that is determined by the effective mass $m^*=0.036(2)~m_\text{e}$, implying that the extent of mass renormalization caused by Kondo hybridization is nearly the same between bulk and nanoflakes. More strikingly, across $T_\text{N1}$, the upturn of FFT amplitude in the nanoflakes are more drastic than in the bulk sample, suggesting a stronger band modulation by AFM1 ordering. Indeed, the obtained $m^*$ is enhanced from $0.037(1)~m_\text{e}$ (above $T_\text{N1}$) to $0.70(7)~m_\text{e}$ (below $T_\text{N1}$), while $\mu_\text{q}$ is increased by 46.5\%; the enhancement ratios of both $m^*$ and $\mu_\text{q}$ are larger than in bulk. This suggests that the scenario of topologically-protected scattering is likely applicable to nanoflakes as well. In particular, $\mu_\text{q}$ at 1.5 K reaches 3158 cm$^2$/Vs, about 31.4\% higher than that of the bulk. \\

\textbf{DISCUSSION} \\

Finally, it is necessary to briefly discuss about the classic transport mobility $\mu$ that can be deduced from a two-band model analysis on the Hall conductivity $\sigma_{xy}$ (seeing Figure S5) \cite{LuoY-WTe2APL2015}. The obtained temperature dependent carrier concentration ($n_\text{e,h}$) and classic transport mobility ($\mu_\text{e,h}$) are shown in Figure S6, where the subscripts e and h are short for electron and hole, respectively. The maximum $\mu_\text{e}$ reaches approximately 3400 cm$^2$/Vs for bulk sample and 2000 cm$^2$/Vs for D3 nanoflake at 1.5 K. This, again, demonstrates the high mobility of CeTe$_3$. An enhancement of $\mu_\text{e}$ is also observed around $T_\text{N1}$, resembling that of $\mu_\text{q}$. We note that the values of $\mu_\text{q}$ and $\mu_\text{e}$ are of the same order of magnitude in CeTe$_3$, hinting that large-angle scattering process is still dominant \cite{AndoT-RMP1982}, despite the non-trivial topological nature. This is in sharp contrast to most non-interacting topological materials such as NbAs where $\mu$ is much larger than $\mu_\text{q}$ due to the suppression of backward scattering by topological protection \cite{LuoY-NbAsSdH}. This counter-indicates that Kondo effect plays an important role in the transport properties of CeTe$_3$ \cite{GorkovLP-CeIn3PRB2006,Haines-CeGeLTP2012}. The high transport mobility in CeTe$_3$ is also reflected by the low scattering rate $1/\tau_1$ determined from fitting optical conductivity to Eq.~(1), seeing Figure S3. An upturn in $1/\tau_1$ is visible below $T_K^{on}$, in line with the onset of Kondo scattering revealed by electrical transport.

In conclusion, we present a comprehensive survey of the vdW AFM Kondo lattice CeTe$_3$ using optical spectroscopy and high-field quantum oscillation measurements. Our results not only complete the Fermi surface topology proposed by earlier ARPES experiment, but also confirm the effective mass enhancement by both AFM1-induced band-structure modulation and Kondo-hybridization-induced narrow band. Despite the mass enhancement, the low-temperature quantum mobility surprisingly increases and reaches $\sim$2400-3200 cm$^2$/Vs. The magnetic orders, high mobility and correlation effect are inherited or even promoted in atomically thin nanoflakes. All these findings suggest that vdW CeTe$_3$ is a rare example that possesses both high charge mobility and electronic correlation, providing a fascinating platform to study 2D Kondo physics and to manipulate novel quantum states. \\

\textbf{METHODS}\\

\textbf{Crystal growth and characterizations}

CeTe$_3$ single crystals were grown using the self-flux method. High-purity raw materials Ce and Te were weighed and mixed in a molar ratio of 3 : 97. The mixture was placed in an alumina crucible, vacuum-sealed in a quartz tube, and heated up to 900 $^\circ$C within 12 hours, held for 12 hours, and then slowly cooled to 550 $^\circ$C at a rate of 2 $^\circ$C/h. After maintaining at 550 $^\circ$C for 6 hours, the excess Te flux was removed by centrifugation. Plate-like single crystals with dimensions up to 3$\times$2$\times$0.5 mm$^3$ were obtained. The obtained single crystals were verified by single-crystal X-ray diffraction using XtaLAB mini II single-crystal diffractometer. This material behaves moderate sensitivity to air and should be stored in an anoxia and dry environment.

\textbf{Thin-flake exfoliation and device fabrication}

Thin flakes of CeTe$_3$ and hBN (hexagonal boron nitride) crystals were mechanically exfoliated onto 290 nm SiO$_2$/Si substrates. hBN and CeTe$_3$ flakes were sequentially picked up using standard van der Waals dry transfer method in a glove box ensuring less than 0.01 ppm O$_2$ and H$_2$O level, and the final stacks of hBN/ CeTe$_3$ were placed directly onto pre-patterned Ti/Au electrodes (5/40 nm) ensuring good Ohmic contacts. hBN and residual polymer introduced in the transfer procedure serve as protection against ambient environment before loading into cryostat. The thickness of CeTe$_3$ flakes were confirmed using atomic force microscopy after transport measurements.

\textbf{Physical properties measurements}

Commercial Magnetic Properties Measurement System (MPMS-VSM, Quantum Design) was employed to measure the magnetic susceptibility. Electrical transport experiments were performed in an IntegraAC Mk II recondensing cryostat equipped with a 16 T superconducting magnet (Oxford Instruments) with a home-built rotation probe. dHvA quantum oscillation measurements under a pulsed magnetic field up to 54 T were carried out at Wuhan National High Magnetic Field Center (WHMFC, China). Quantum oscillations of bulk CeTe$_3$ were also studied by SdH effect via magnetoresistivity measurements for field up to 34 T at static High Magnetic Field Laboratory (Hefei, China). For these quantum oscillation measurements, the internal field was corrected by calculating the demagnetization factor \cite{AharoniA-JAP1998}.

\textbf{Optical conductivity measurements}

The reflectivity $R(\omega)$ of CeTe$_3$ was measured using a Bruker Vertex 80v Fourier transform spectrometer, with the light polarized in the $\mathbf{ac}$ planes. An in-situ gold overfilling technique was used to obtain the absolute $R(\omega)$ at different temperatures between 5 and 300 K over a broad frequency ranging from 50 to 7000 cm$^{-1}$. $R(\omega)$ from 7000 to 25000 cm$^{-1}$ was also measured at room temperature. Then, we employed an AvaSpec-2048$\times$14 optical fiber spectrometer to measure $R(\omega)$ in the frequency up to 50000 cm$^{-1}$ at room temperature. The real part of the optical conductivity $\sigma_1(\omega)$ was determined from a Kramers-Kronig analysis of the measured $R(\omega)$. We adopted a Hagen-Rubens ($R=1-A\sqrt{\omega}$) low-frequency extrapolation below 50 cm$^{-1}$. A constant $R(\omega)$ up to 12.5 eV, followed by a free-electron ($\omega^{-4}$) response, was used for the high-frequency extrapolation. For comparison, similar measurements were also performed on LaTe$_3$. \\

\textbf{RESOURCE AVAILABILITY} \\

\textbf{Lead contact}

Further information and requests for resources or other information should be directed to and will be fulfilled by the lead contact, Yongkang Luo (mpzslyk@gmail.com)

\textbf{Materials availability}

This study did not generate new unique materials.

\textbf{Data and code availability}

The raw data that support the findings of this study are available from the corresponding authors upon proper request. \\

\textbf{ACKNOWLEDGMENTS} \\

This research was funded by the National Key R\&D Program of China (2023YFA1609600 and 2022YFA1602602), the National Natural Science Foundation of China (U23A20580, 12174180 and 12274155), the Guangdong Basic and Applied Basic Research Foundation (2022B1515120020), and the Beijing National Laboratory for Condensed Matter Physics (2024BNLCMPKF004). We thank the HM (https://cstr.cn/31125.02.SHMFF.HM) at the Steady High Magnetic Field Facility, CAS (https://cstr.cn/31125.02.SHMFF), for providing technical support and assistance in data collection and analysis.\\

\textbf{AUTHOR CONTRIBUTIONS}\\

Y.L. conceived and designed the experiments. H.Z. grew the crystals and performed most of the bulk-property measurements with the aids from S.Z., Z.W., K.L. and Y.Y. C.D. and J.W. carried out the dHvA measurements with pulsed-field magnetization facility with the technical support of L.L. J.C., J.Z., C.X. measured the SdH quantum oscillations under static high magnetic field. B.J. and Y.D. made the optical conductivity measurements. Y.Z. and J.L. fabricated the nanoflakes and performed the SdH measurements on the devices. K.W. provided constructive suggestions. H.Z., Y.D., J.L. and Y.L. discussed the data, interpreted the results, and wrote the paper with input from all the authors. \\

\textbf{DECLARATION OF INTERESTS} \\

The authors declare no competing interests. \\

\textbf{SUPPLEMENTAL INFORMATION} \\

Supplemental information can be found online at ********  \\


\clearpage

\renewcommand{\thefigure}{S\arabic{figure}}
\renewcommand{\thetable}{S\arabic{table}}
\renewcommand{\theequation}{S\arabic{equation}}
\setcounter{table}{0}
\setcounter{figure}{0}
\setcounter{equation}{0}
\setcounter{page}{1}

\vspace{-15pt}

\begin{center}
\large
\textbf{Supplemental Information: } \\
\textbf{Kondo-coupled van der Waals antiferromagnet with high-mobility quasiparticles}\\
\small
\emph{}\\
Hai Zeng$^{1*}$, Yang Zhang$^{1*}$, Bingke Ji$^{2*}$, Jiaqiang Cai$^{3}$, Shuo Zou$^{1}$, Zhuo
Wang$^{1}$, Chao Dong$^{1}$, Kangjian Luo$^{1}$, Yang Yuan$^{1}$, Kai Wang$^{4}$, Jinglei Zhang$^{3}$, Chuanyin
Xi$^{3}$, Junfeng Wang$^{1}$, Liang Li$^{1,5}$, Yaomin Dai$^{2\dag}$, Jing Li$^{1\ddag}$, and Yongkang Luo$^{1,5,6\S}$\\
$^1$ {\it Wuhan National High Magnetic Field Center and School of Physics, Huazhong University of Science and Technology, Wuhan 430074, China;} \\
$^2$ {\it School of Physics, Nanjing University, Nanjing 210023, China;} \\
$^3$ {\it Anhui Key Laboratory of Low-Energy Quantum Materials and Devices, High Magnetic Field Laboratory, Hefei Institutes of Physical Science, Chinese Academy of Sciences, Hefei 230031, China;}\\
$^4$ {\it School of Physics and Electronics, Henan University, Kaifeng 475004, China;}\\
$^5$ {\it State Key Laboratory of Advanced Electromagnetic Technology, Huazhong University of Science and Technology, Wuhan 430074, China;}\\
$^6$ {\it Lead contact}\\
\end{center}

$^{*}$ These authors contribute equally to this work

$^{\dag}$ Corresponding author: Yaomin Dai (ymdai@nju.edu.cn)

$^{\ddag}$ Corresponding author: Jing Li (jing\_li@hust.edu.cn)

$^{\S}$ Corresponding author: Yongkang Luo (mpzslyk@gmail.com) \\

\normalsize

\emph{} \\

In this \textbf{Supplemental Information} (\textbf{SI}), we provide additional results that will further support the discussions and conclusion in the main text, including crystalline characterization, magnetization, magnetoresistance, optical spectroscopy, Hall effect and two-band fitting, and Lifshitz-Kosevich analysis of quantum oscillations.

\newpage

\clearpage
\textbf{SI \Rmnum{1}: XRD, magnetization, and magnetoresistance}\\

\begin{figure*}[!htp]
	\vspace*{-0pt}
	\includegraphics[width=15.3cm]{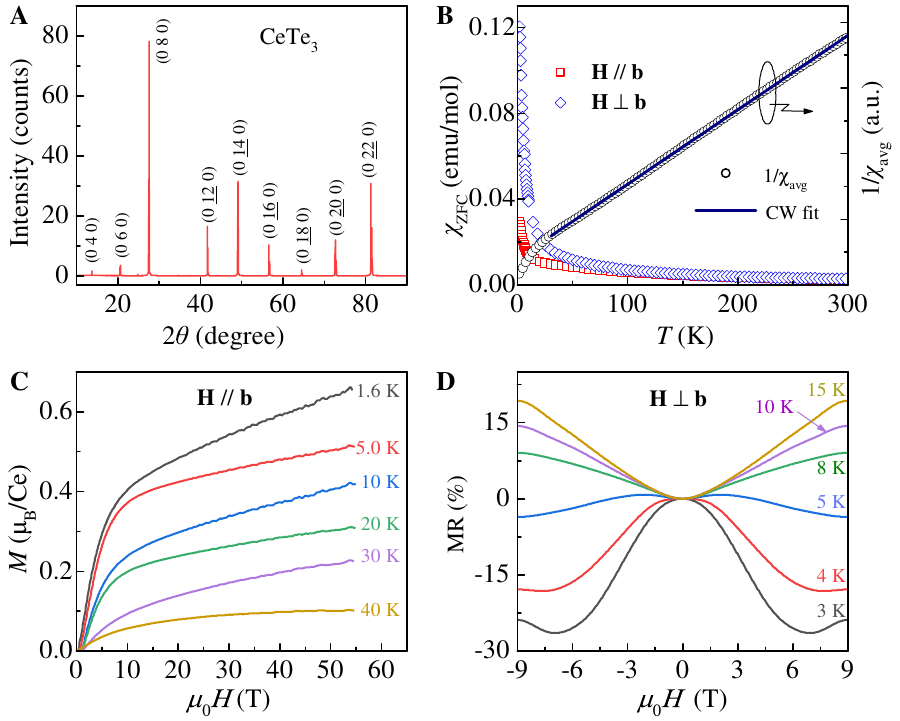}
	\label{Fig.S1}
\vspace*{0pt}
\small
\begin{flushleft}
\justifying{
    \textbf{Figure S1. XRD, magnetization, and magnetoresistance.}\\ (A) X-ray diffraction pattern of a CeTe$_3$ single crystal, where only $(0, 2k, 0)$ diffractions are observed. The lattice parameters $a$, $b$ and $c$ at room temperature are 4.24(9), 23.25(4) and 4.52(3) \AA, respectively. The large ratio $b/a(c) \sim 5.5$ indicates a quasi-two-dimensional character of the crystal structure.\\ (B) Temperature-dependent magnetic susceptibility of bulk CeTe$_3$ for different crystallography directions. The navy solid line represents the Curie-Weiss fit of the inverse powder-averaged magnetic susceptibility, $1/\chi_\text{avg}=3/(\chi_b+2\chi_{ac})$. The fitting yields effective magnetic moment $\mu_\text{eff}$ = 2.62 $\mu_\text{B}$ and the Weiss temperature $\theta_\text{W} = -35.11$ K.\\ (C) Isothermal field-dependent magnetization curves under pulsed high magnetic field along $\mathbf{b}$. dHvA quantum oscillations can be observed at high field.\\ (D) In-plane magnetoresistance (MR) of bulk sample. The curve exhibits negative MR at low temperature and gradually turns into positive with increasing temperature.
    }
\end{flushleft}
\normalsize
\end{figure*}

\newpage

\textbf{SI \Rmnum{2}:  Optical spectroscopy data of CeTe$_3$ and LaTe$_3$}\\

\begin{figure*}[!htp]
	\vspace*{-10pt}
	\includegraphics[width=15.3cm]{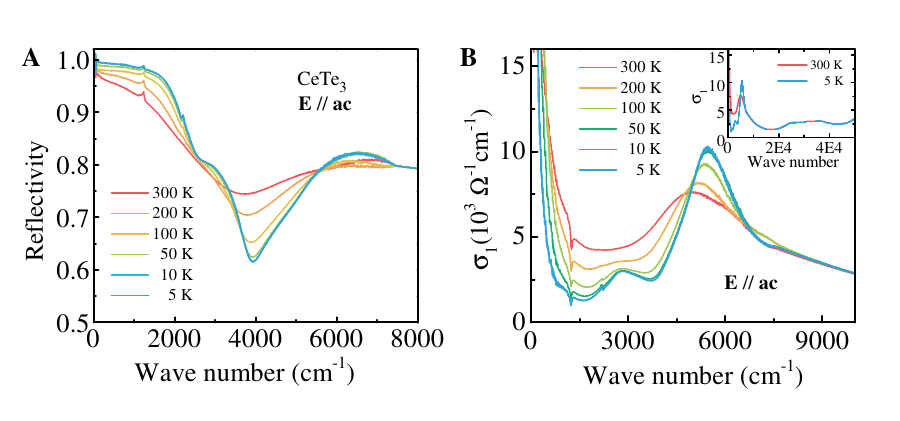}
	\label{Fig.S2}
\vspace*{-10pt}
\small
\begin{flushleft}
\justifying{
    \textbf{Figure S2. Optical spectroscopy data of CeTe$_3$.}\\ (A) Reflectivity spectra $R(\omega)$ up to 8000 cm$^{-1}$ at different temperatures.\\ (B) Real part of the optical conductivity, $\sigma_1(\omega)$. Signatures for multiple CDW gaps are observed near 5000 cm$^{-1}$ for all the measured temperatures and 2800 cm$^{-1}$ for below 200 K. Inset shows $\sigma_1(\omega)$ at representative 300 and 5 K in the full spectral range 50 - 50000 cm$^{-1}$.
    }
\end{flushleft}
\normalsize
\end{figure*}

\begin{figure*}[!htp]
	\vspace*{-10pt}
	\includegraphics[width=10cm]{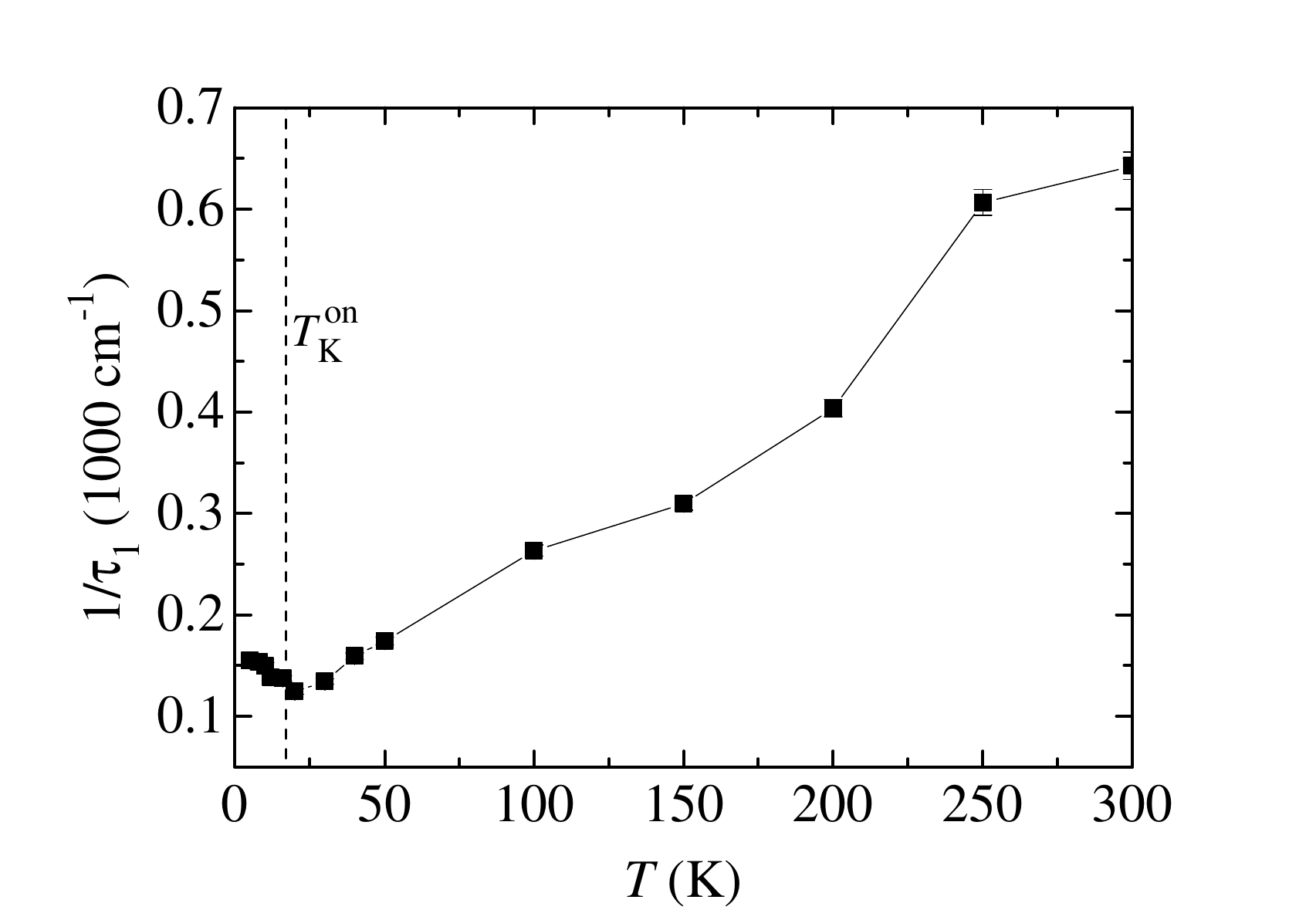}
	\label{Fig.S3}
\vspace*{-0pt}
\small
\begin{flushleft}
\justifying{
    \textbf{Figure S3. Temperature dependent $1/\tau_1$ extracted from optical conductivity.} \\ On the whole, $1/\tau_1$ is very small, characteristic of low scattering rate and high charge mobility. An upturn in $1/\tau_1$ is observed near $T_\text{K}^{\text{on}}$ due to the onset of incoherent Kondo scattering. The drop of $1/\tau_1$ near 200 K is likely owing to the CDW order.
    }
\end{flushleft}
\normalsize
\end{figure*}

\newpage

\begin{figure*}[!htp]
	\vspace*{-0pt}
	\includegraphics[width=15.3cm]{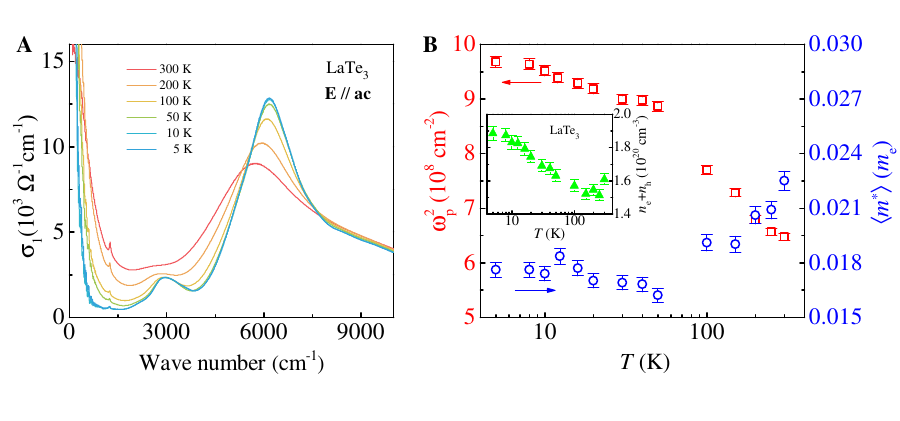}
	\label{Fig.S4}
\vspace*{-15pt}
\small
\begin{flushleft}
\justifying{
    \textbf{Figure S4. Optical spectroscopy data of LaTe$_3$.}\\ (A) Real part of the optical conductivity, $\sigma_1(\omega)$. Signatures for multiple CDW gaps are observed near 6000 cm$^{-1}$ and 3000 cm$^{-1}$. \\ (B) Left, temperature dependence of $\omega_\text{p}^2\equiv\omega_\text{p1}^2+\omega_\text{p2}^2$; right, temperature dependent average effective mass of charge carriers, $\langle m^{*} \rangle$. The inset shows the sum of hole and electron carrier concentrations. Note that upon cooling, $\omega_\text{p}^2$ increases monotonically at low temperature, which stands in stark contrast to that of CeTe$_3$. This leads to the great difference in $\langle m^* \rangle$ between LaTe$_3$ and CeTe$_3$.
    }
\end{flushleft}
\normalsize
\end{figure*}

\newpage

\textbf{SI \Rmnum{3}: Magnetotransport of CeTe$_3$}\\

\begin{figure*}[!htp]
	\vspace*{-0pt}
	\includegraphics[width=15cm]{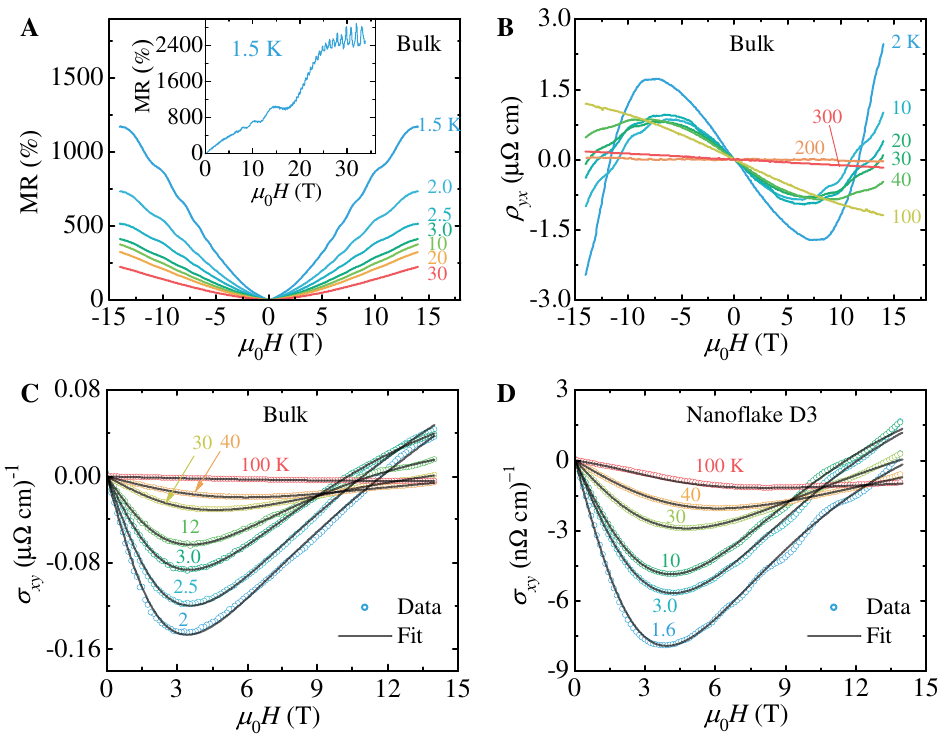}
	\label{Fig.S5}
\vspace*{-0pt}
\small
\begin{flushleft}
\justifying{
    \textbf{Figure S5. Magnetotransport data of CeTe$_3$.}\\ (A) $\mathbf{H}\parallel\mathbf{b}$ magnetoresistance (MR) of bulk CeTe$_3$. The inset shows the MR curve at 1.5 K up to 34 T.\\ (B) Hall resistivity $\rho_{yx}$ of bulk sample at selected temperatures.\\ (C) Hall conductivity $\sigma_{xy}$ of bulk sample, and fitting to a two-band model.\\ (D) Hall conductivity $\sigma_{xy}$ of nanoflake sample D3, and fitting to a two-band model.
    }
\end{flushleft}
\normalsize
\end{figure*}

\begin{figure*}[!htp]
	\vspace*{-0pt}
	\includegraphics[width=15.3cm]{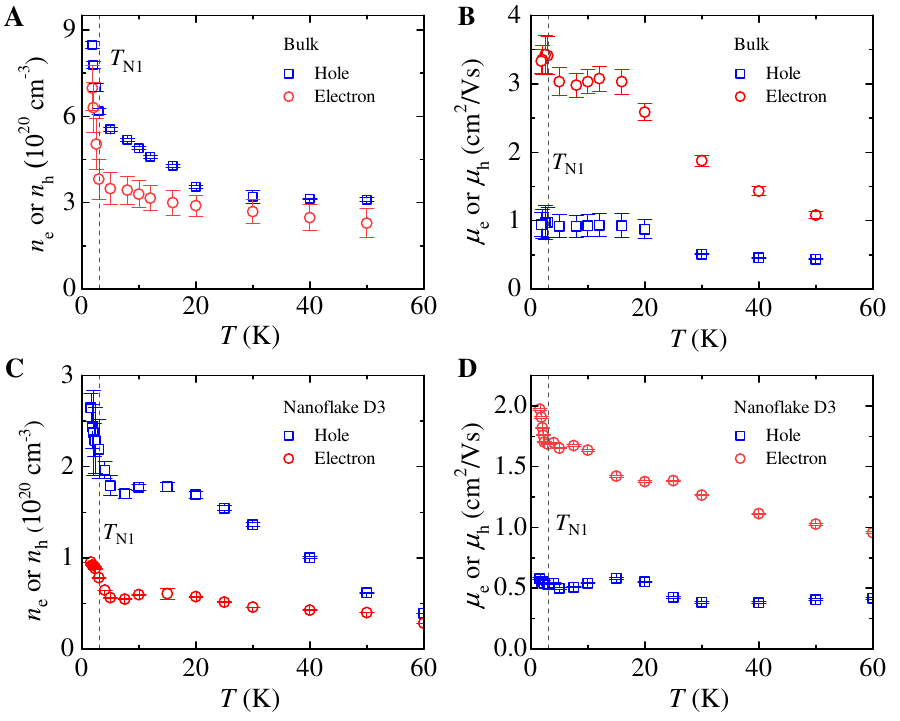}
	\label{Fig.S6}
\vspace*{-0pt}
\small
\begin{flushleft}
\justifying{
    \textbf{Figure S6. Carrier density and transport mobility of bulk and nanoflake D3 samples.}\\ (A) $n_\text{e,h}$ of bulk CeTe$_3$.\\ (B) $\mu_\text{e,h}$ of bulk CeTe$_3$.\\ (C) $n_\text{e,h}$ of CeTe$_3$ nanoflake D3.\\ (D) $\mu_\text{e,h}$ of CeTe$_3$ nanoflake D3.
    }
\end{flushleft}
\normalsize
\end{figure*}

\clearpage

\textbf{SI \Rmnum{4}: Additional quantum-oscillation data of CeTe$_3$}\\

\begin{figure*}[!htp]
	\vspace*{-0pt}
	\includegraphics[width=15cm]{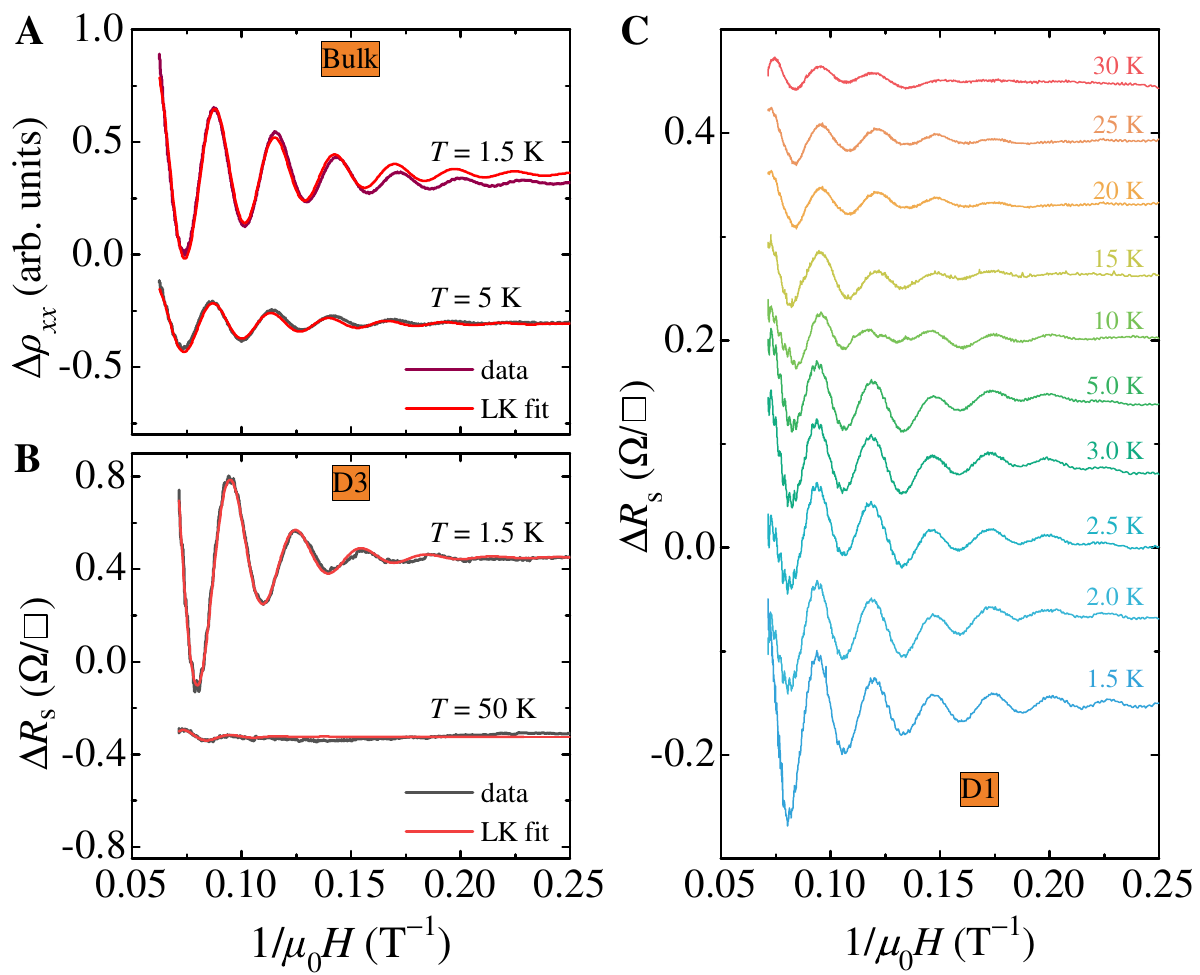}
	\label{Fig.S7}
\vspace*{-0pt}
\small
\begin{flushleft}
\justifying{
    \textbf{Figure S7. Additional SdH oscillation data of CeTe$_3$.}\\ (A) Raw data and fitting to LK formula for bulk sample at 1.5 K (AFM regime) and 5 K (PM regime).\\ (B) Results for D3, at 1.5 K and 50 K.\\ (C) SdH oscillations of D1 at different temperatures ranging from 1.5 to 30 K.
    }
\end{flushleft}
\normalsize
\end{figure*}

\begin{figure*}[!htp]
	\vspace*{-0pt}
    \hspace*{-10pt}
	\includegraphics[width=17.0cm]{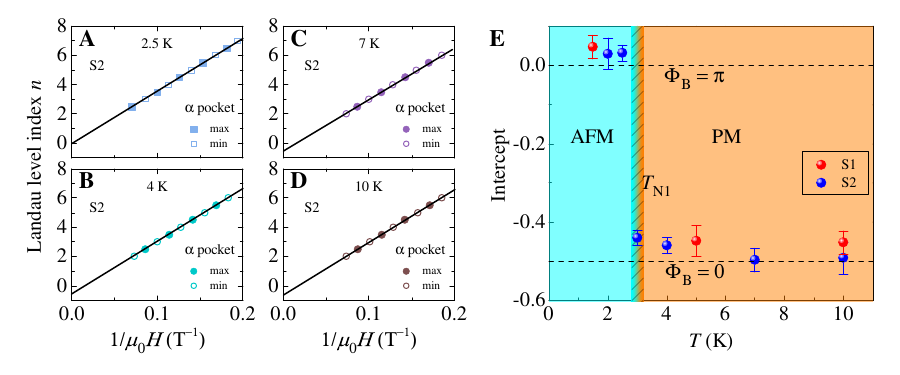}
	\label{Fig.S8}
\vspace*{-20pt}
\small
\begin{flushleft}
\justifying{
    \textbf{Figure S8. Additional evidence for the topological transition across $T_{\text{N}1}$ in CeTe$_3$.}\\ (A-D) Landau fan index diagram of sample S2 at selected temperatures.\\ (E) Temperature dependent intercept of the Landau fan plot. The hatched area indicates the AFM1 transition near $T_{\text{N}1}$. A jump of Berry phase $\Phi_\text{B}$ from $\pi$ to $0$ is clearly seen near $T_{\text{N}1}$. Sample S2 was measured under field up to 16 T, while sample S1 up to 34 T (shown in Figure 3).
    }
\end{flushleft}
\normalsize
\end{figure*}

\end{document}